\documentclass[%
 aip,
 amsmath,amssymb,
 reprint,%
]{revtex4-1}

\usepackage{graphicx}
\usepackage{dcolumn}
\usepackage{bm}

\usepackage{siunitx}
\usepackage{appendix}
\usepackage{amsmath}
\usepackage{amssymb}
\usepackage{soul}
\usepackage{physics}
\usepackage{xcolor}

\usepackage[utf8]{inputenc}
\usepackage[T1]{fontenc}
\usepackage{mathptmx}
\usepackage{etoolbox}

\makeatletter
\def\@email#1#2{%
 \endgroup
 \patchcmd{\titleblock@produce}
  {\frontmatter@RRAPformat}
  {\frontmatter@RRAPformat{\produce@RRAP{*#1\href{mailto:#2}{#2}}}\frontmatter@RRAPformat}
  {}{}
}%
\makeatother
\begin{document}

\preprint{AIP/123-QED}

\title[Deployed entanglement-based BBM92 quantum key distribution using frequency-converted photons emitted by a
quantum dot]{Deployed entanglement-based BBM92 quantum key distribution using frequency-converted photons emitted by a
quantum dot}

\author{Michal~Vyvlecka}
\email{michal.vyvlecka@ihfg.uni-stuttgart.de}
\altaffiliation[These authors contributed equally: Michal Vyvlecka, Raphael Joos. Authors to whom correspondence should be addressed: michal.vyvlecka@ihfg.uni-stuttgart.de and r.joos@ihfg.uni-stuttgart.de]{}
\affiliation{Institut f\"ur Halbleiteroptik und Funktionelle Grenzfl\"achen, \\Center for Integrated Quantum Science and Technology (IQ\textsuperscript{ST}) and SCoPE, \\University of Stuttgart, Allmandring 3, 70569 Stuttgart, Germany}%

\author{Raphael~Joos}
\email{r.joos@ihfg.uni-stuttgart.de}
\altaffiliation[These authors contributed equally: Michal Vyvlecka, Raphael Joos. Authors to whom correspondence should be addressed: michal.vyvlecka@ihfg.uni-stuttgart.de and r.joos@ihfg.uni-stuttgart.de]{}
\affiliation{Institut f\"ur Halbleiteroptik und Funktionelle Grenzfl\"achen, \\Center for Integrated Quantum Science and Technology (IQ\textsuperscript{ST}) and SCoPE, \\University of Stuttgart, Allmandring 3, 70569 Stuttgart, Germany}%

\author{Benjamin~Breiholz}
\affiliation{Institut f\"ur Halbleiteroptik und Funktionelle Grenzfl\"achen, \\Center for Integrated Quantum Science and Technology (IQ\textsuperscript{ST}) and SCoPE, \\University of Stuttgart, Allmandring 3, 70569 Stuttgart, Germany}%

\author{Emma~Marmasse}
\affiliation{Institut f\"ur Halbleiteroptik und Funktionelle Grenzfl\"achen, \\Center for Integrated Quantum Science and Technology (IQ\textsuperscript{ST}) and SCoPE, \\University of Stuttgart, Allmandring 3, 70569 Stuttgart, Germany}%

\author{Ilenia~Neureuther}
\affiliation{Institut f\"ur Halbleiteroptik und Funktionelle Grenzfl\"achen, \\Center for Integrated Quantum Science and Technology (IQ\textsuperscript{ST}) and SCoPE, \\University of Stuttgart, Allmandring 3, 70569 Stuttgart, Germany}%

\author{Timo~Schnieber}
\affiliation{Institut f\"ur Halbleiteroptik und Funktionelle Grenzfl\"achen, \\Center for Integrated Quantum Science and Technology (IQ\textsuperscript{ST}) and SCoPE, \\University of Stuttgart, Allmandring 3, 70569 Stuttgart, Germany}%

\author{Anna~Frederike~K\"ohler}
\affiliation{Institut f\"ur Halbleiteroptik und Funktionelle Grenzfl\"achen, \\Center for Integrated Quantum Science and Technology (IQ\textsuperscript{ST}) and SCoPE, \\University of Stuttgart, Allmandring 3, 70569 Stuttgart, Germany}%

\author{Tim~Strobel}
\affiliation{Institut f\"ur Halbleiteroptik und Funktionelle Grenzfl\"achen, \\Center for Integrated Quantum Science and Technology (IQ\textsuperscript{ST}) and SCoPE, \\University of Stuttgart, Allmandring 3, 70569 Stuttgart, Germany}%

\author{Andreas~Berg}
\affiliation{TIK – Technical Information and Communication Services, \\University of Stuttgart, Allmandring 30a, 70569 Stuttgart, Germany}%

\author{Sebastian~Kiesel}
\affiliation{TIK – Technical Information and Communication Services, \\University of Stuttgart, Allmandring 30a, 70569 Stuttgart, Germany}%

\author{Tobias~Bauer}
\affiliation{Fachrichtung Physik, Universit\"at des Saarlandes, Campus E2.6, 66123 Saarbr\"ucken, Germany}%

\author{Marlon~Sch\"afer}
\affiliation{Fachrichtung Physik, Universit\"at des Saarlandes, Campus E2.6, 66123 Saarbr\"ucken, Germany}%

\author{Nand~Lal~Sharma}
\affiliation{Institute for Integrative Nanosciences, Leibniz IFW Dresden, Helmholtzstraße 20, 01069 Dresden, Germany}%

\author{Caspar~Hopfmann}
\affiliation{Institute for Integrative Nanosciences, Leibniz IFW Dresden, Helmholtzstraße 20, 01069 Dresden, Germany}%

\author{Simone~Luca~Portalupi}
\affiliation{Institut f\"ur Halbleiteroptik und Funktionelle Grenzfl\"achen, \\Center for Integrated Quantum Science and Technology (IQ\textsuperscript{ST}) and SCoPE, \\University of Stuttgart, Allmandring 3, 70569 Stuttgart, Germany}%

\author{Christoph~Becher}
\affiliation{Fachrichtung Physik, Universit\"at des Saarlandes, Campus E2.6, 66123 Saarbr\"ucken, Germany}%

\author{Peter~Michler}
\affiliation{Institut f\"ur Halbleiteroptik und Funktionelle Grenzfl\"achen, \\Center for Integrated Quantum Science and Technology (IQ\textsuperscript{ST}) and SCoPE, \\University of Stuttgart, Allmandring 3, 70569 Stuttgart, Germany}%

\date{\today}

\begin{abstract}
Establishing secure communication is an important requirement in existing and future networks. Although current solutions rely on classical approaches, quantum mechanics offers the possibility of moving from computational to information-theoretic security. Quantum key distribution (QKD) has been extensively studied as a means of achieving such security. Recently, epitaxial quantum dots started to be studied and employed in quantum cryptography, because of their rapid development that made them one of the brightest sources of quantum light to date. In particular, they are known to be excellent sources of polarization-entangled photon pairs, especially droplet-etched dots operating at $\sim$\SI{780}{\nano\meter}. Here, we employ such a quantum dot to generate entangled photon pairs under \SI{1}{\giga\hertz} excitation rate. Using quantum frequency conversion, we interface the emitted photons with the telecommunication wavelength range while preserving their polarization entanglement. This enables quantum key distribution over a \SI{1.59}{\kilo\meter} field-deployed fibre link across the University of Stuttgart - campus Vaihingen, using the BBM92 protocol: we achieve a secure key rate of \SI{206}{\hertz} over \SI{19}{\hour} of stable operation, resulting in a total secure key length of \SI{14.2}{\mega\bit}. This key enables the safe encryption and decryption of a colour image with further implementation of error correction and privacy amplification. These results demonstrate important steps towards hybrid quantum networks by combining high-quality entangled-photon sources operated at high excitation rates, efficient frequency conversion module, and picosecond-level time synchronization in a realistic deployed network.

\end{abstract}

\maketitle

Quantum cryptography employs laws of quantum theory, such as the no-cloning theorem and the uncertainty principle, to design cryptographic protocols that guarantee security against an adversary with unlimited computational power~\cite{Bennett1984,Bozzio2025}. Over the last decade, many quantum cryptographic primitives, basic building blocks of more complex quantum cryptographic protocols, have been theoretically described and experimentally demonstrated~\cite{Pan:RevMod20,NJC:NatComms12,Pan:PRL14,GAA:pra18,BCK:PRA20}. The most mature one, quantum key distribution, has been many times implemented on deployed optical fiber networks and is almost routinely used in scenarios which require highly secure communication~\cite{Pan:RevMod20,Peev2009,Chen_npj2021}. These advances are the main driving forces for realizing a universal long-distance quantum communication network, so-called quantum internet~\cite{Kimble:2008,WEH:Sci18}, which allows implementation of quantum cryptographic protocols in real-life scenarios.

These days, most deployed quantum cryptographic implementations are based on photons emitted by Poisson-distributed sources (PDSs), such as spontaneous parametric down-conversion sources and attenuated lasers~\cite{Pan:RevMod20,Bozzio2025}. These single-photon sources demonstrate high technological readiness, robustness to environmental fluctuations, room-temperature operation, and low-cost, high-yield  production. This makes them the leading sources in modern quantum cryptography, despite not being ideal single-photon sources. The main limitation of these sources comes from the trade-off between the multi-photon contribution and the source brightness, which limits their performance at long-distance quantum communication~\cite{SKB:JO19}, lowering the efficiency of possible quantum repeaters and making the implementation of mistrustful quantum cryptographic protocols challenging~\cite{BLM:PRL00,Bozzio2025}.

\begin{figure*}
	\begin{center}

		\includegraphics[width=\linewidth]{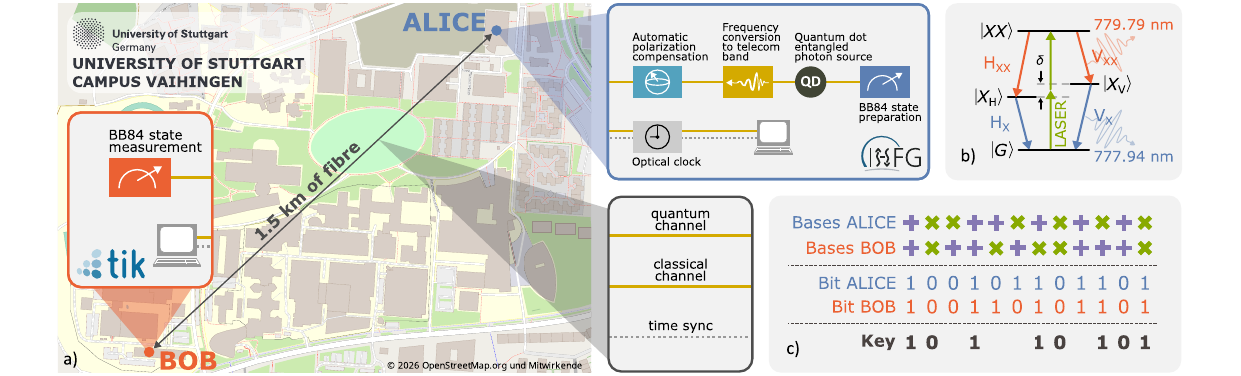}
		\caption{\textbf{Experimental setup and measurement configuration.} a) Map of the Vaihingen campus of the University of Stuttgart, showing the two experimental locations: Alice, at the IHFG (Allmandring 3), and Bob, at the TIK (Allmandring 30), connected by \SI{1.59}{\kilo\meter} of deployed optical fibre. At the Alice location (see inset), the source configuration is schematically illustrated: the quantum dot emits entangled photon pairs, one of which is sent to the BB84 state-preparation unit, while the second photon is first frequency-converted to the telecom C-band and subsequently transmitted through the deployed fibre link. An automated polarization-compensation unit provides active control of the polarization state of the deployed fiber link. An optical clock provides the time reference for synchronization between the two locations. Bob (see inset) contains the BB84 state-measurement setup. The deployed fibre link (see insert comprises three fibers: one for the quantum channel, one for classical communication, and one for optical time synchronization. b) Schematic of the key-sifting procedure between Alice and Bob, measurement outcomes are retained when Alice and Bob select the same measurement basis and discarded otherwise. c) Energy-level diagram of the biexciton--exciton--ground-state cascade in the semiconductor quantum dot generating two entangled photons with a spectral separation of \SI{1.85}{\nano\meter}. 
        }
		\label{fig:fig1}
	\end{center}
\end{figure*}

This limitation can be overcome by employing quantum dot single-photon sources (QDs) in quantum cryptographic protocols, as QDs have demonstrated both high brightness and high single-photon purity~\cite{Michler2017,Ding2025}. Moreover, multiple excitation schemes allow QDs to emit photons without photon-number coherence~\cite{Bozzio2022,Karli2024,Vyvlecka2023} or entangled photon pairs with high fidelity~\cite{Mueller2014,Huber2017}. The combination of these tunable parameters makes QDs promising candidates as quantum light sources for quantum cryptography applications, with the potential to surpass the performance of currently used PDSs~\cite{Bozzio2022,Zhang2025}. Indeed, quantum dots have been employed in a variety of quantum communication protocols, ranging from deployed QKD employing single-~\cite{Yang2024,Zahidy2024} or entangled photons~\cite{Schimpf2021,BassoBasset2021,Rota2026}, to quantum random number generation~\cite{Gyger2022}, decoy-state protocols over long fibres~\cite{BrookeBarnes2026}, and, more recently, strong coin flipping~\cite{Vajner2026}.

The future global quantum internet will comprise numerous building blocks, including quantum memories, quantum repeaters, and interfaces to quantum computing nodes~\cite{Wei2022}. These components may rely on a variety of physical platforms, such as quantum dots~\cite{Yu2023}, spontaneous parametric down-conversion (SPDC) sources~\cite{Ma2012,Ren2017}, trapped atoms~\cite{Ritter2012}, nitrogen-vacancy centers~\cite{Hensen2015}, or other quantum systems. Interconnecting these heterogeneous platforms into a unified quantum network remains one of the major scientific and technological challenges in developing the quantum internet~\cite{Awschalom2021}.

To establish photonic interconnections between different quantum platforms, the wavelengths of the exchanged photons must be matched~\cite{Awschalom2021}. In contrast to photons generated by spontaneous parametric down-conversion sources~\cite{Couteau2018}, the emission wavelength of a quantum dot can only be tuned over a relatively limited spectral range~\cite{Patel2010,Martin-Sanchez2018,Yang2024}. Quantum frequency converters represent an appealing approach to bridge the wavelength mismatch between different quantum platforms and to match the target quantum system's acceptance wavelength~\cite{Bock2018,vanLeent2022,Arensktter2023,Weber2019}.

In this work, we combine a droplet-etched epitaxially grown GaAs quantum dot with a low fine-structure splitting (FSS), emitting highly entangled photon pairs at approximately \SI{780}{\nano\meter}, with a polarization-maintaining quantum frequency converter (QFC) that converts one photon of each entangled pair to \SI{1515}{\nano\meter} close to the telecommunication C-band. This wavelength conversion makes the photon suitable for long-distance quantum communication over standard single-mode silica optical fibres while preserving its entanglement~\cite{Strobel2024}.

To demonstrate the performance of this hybrid quantum interface, combining a high-performance entangled-photon source with a QFC, we implemented the entanglement-based quantum key distribution (QKD) protocol BBM92~\cite{Bennett1992} over an existing deployed fibre network on the Vaihingen campus of the University of Stuttgart. The communicating parties, Alice and Bob, were physically separated by approximately \SI{700}{\meter} as the crow flies and connected through \SI{1.59}{\kilo\meter} of deployed optical fibre with total losses of \SI{1.09}{\decibel}, for characterization of the deployed fibre link, see the Supplementary Material. Alice's setup consisted of standard laboratory-based optical components located in the stable laboratory environment of the Institute of Semiconductor Optics and Functional Interfaces (IHFG), whereas Bob's setup, including the projective measurement apparatus, control electronics, and single-photon detectors, was integrated into a standard 19-inch rack and placed in an office space of the Technical Information and Communication Services (TIK) department of the University of Stuttgart.

The layout of the Stuttgart-Vaihingen campus quantum network testbed used for the entanglement-based QKD implementation is shown in Fig.~\ref{fig:fig1} a). The first communication party, Alice, comprises an optically pumped GaAs QD that emits polarization-entangled photon pairs under pulsed resonant two-photon excitation (TPE). 
To generate entangled photon pairs, we implemented a GaAs QD obtained by droplet epitaxy, embedded in a broadband optical antenna design to enhance photon-pair extraction and fibre-coupling efficiencies over bulk samples~\cite{Strobel2024}. The entangled photon-pair state generated by the biexciton-exciton cascade decay has the form
$\ket{\psi(t)}=\frac{1}{\sqrt{2}}\left(\ket{H_{XX}H_X}+e^{-i\delta t/\hbar}\ket{V_{XX}V_X}\right)$,
where $\delta$ is the FSS, which causes the relative phase between the two components of the entangled state to evolve with the time difference between the emissions of the XX and X photons. The QD used in this work exhibits an FSS of $\delta$=\SI{2.1\pm0.3}{\micro\eV} which, for a coincidence time window of \SI{275}{\pico\second}, results in a state close to the Bell state $\ket{\phi^+}$ with a fidelity of \SI{91.4}{\percent}, the corresponding density matrix is shown in the Supplementary Material. 

The QD was excited via a two-photon excitation (TPE) scheme at \SI{779}{\nano\meter}. For this, second-harmonic generation of telecom C-band laser pulses by a fibre-coupled periodically poled lithium niobate (PPLN) waveguide was used. The fundamental wave was generated by a fibre-based pulsed laser, emitting optical pulses with a central wavelength of \SI{1557.73}{\nano\meter}, a pulse width of \SI{10}{\pico\second}, corresponding to a spectral FWHM of \SI{400}{\pico\meter}, and operating at a repetition rate of \SI{1}{\giga\hertz}. This excitation setup was chosen to enable a high repetition rate with stable pulse-to-pulse operation within a compact footprint and without requiring additional repetition-rate multipliers. 
The biexciton photon (XX), emitted at a wavelength of \SI{779.79}{\nano\meter}, is frequency-converted to \SI{1515}{\nano\meter} using a QFC module based on a Sagnac interferometer~\cite{Bock2018,vanLeent2022}. The Sagnac configuration preserves the polarization state and, consequently, the entanglement during the conversion process. The QFC module achieves a conversion efficiency of up to \SI{50}{\percent}. The converted XX photon, now at a wavelength suitable for long-distance transmission through standard single-mode silica optical fibre, is transmitted through the deployed fibre link to the second communication party, Bob, located on the opposite side of the university campus.

Polarization drifts in the deployed fibre link, caused by environmental fluctuations such as temperature variations over the day, are compensated using an automated polarization-control unit based on multiple motorized wave plates. The compensation algorithm probes the fibre link using well-defined input polarization states encoded using a CW alignment laser and optimizes the corresponding Stokes parameters by adjusting the wave plates in the polarization-control module.

The temporal synchronization between Alice and Bob is established by sharing an optical frequency reference distributed through the fibre network and generated by an ultrafast optical clock. Classical communication between the two parties is performed over the university's local area network (LAN). A detailed description of the individual modules of the quantum network testbed is provided in the Supplementary Material.

The implemented QKD protocol is the entanglement-based BBM92 protocol, in which the BB84 states are established through remote state preparation. Before the XX photon is transmitted to Bob, the  exciton photon (X), is measured by Alice in one of two mutually unbiased conjugate bases, here chosen as the (Z)- and (X)-bases. The choice of measurement basis is implemented passively using only a 50:50 beam splitter, thereby limiting the communication rate primarily by the photon-detection jitter and the QD excitation repetition rate, and not by the speed of active elements needed in non-entanglement based protocols. The measurement of the X photon projects the XX photon of the entangled pair onto the corresponding polarization state determined by Alice's measurement outcome. Consequently, Alice and Bob share correlated states from the set ($\{H,V,D,A\}$), corresponding to horizontal, vertical, diagonal, and antidiagonal polarization, respectively. These four states constitute the BB84 state set used for the subsequent QKD protocol, with Bob performing an analogous measurement on the received XX photon, randomly selecting between the (Z)- and (X)-bases. 

When Alice and Bob choose the same measurement basis, their measurement outcomes are ideally correlated. In practice, limited fidelity of the entangled state, fluctuations of transmission channel, and imperfections of measurement systems result in a finite error probability, so-called quantum bit error rate (QBER). Events for which Alice and Bob selected the same basis are therefore retained during the sifting phase of the protocol, and their corresponding measurement outcomes form the raw key, as illustrated in Fig.~\ref{fig:fig1}b).

\begin{figure}
	\begin{center}
    
		\includegraphics[width=\linewidth]{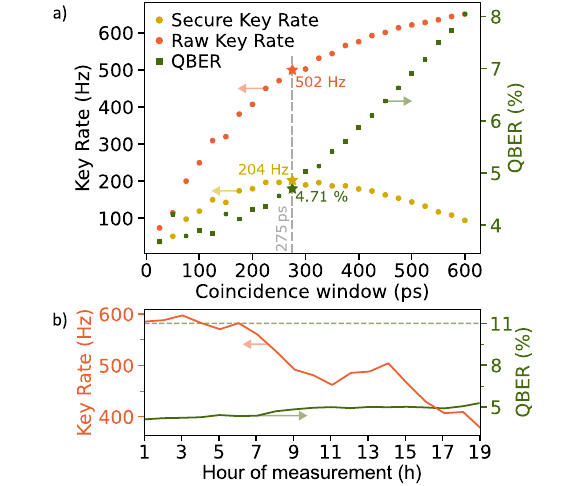}
		\caption{\textbf{Key experimental figures: Key rate, secure key rate, and QBER.} a) Observed raw key rate (RKR) (left scale) and QBER (right scale) as a function of the coincidence window used for the detection of the two photons. For a coincidence window of \SI{275}{\pico\second}, the secure key rate (SKR) reaches its maximum of \SI{204}{\hertz}, with an RKR of \SI{502}{\hertz}, while maintaining a QBER of \SI{4.71}{\percent}, well below the theoretical security threshold of \SI{11}{\percent} for the present protocol. b) Raw key rate (left scale) and QBER (right scale) as a function of time during the \SI{19}{\hour}-long measurement. The plotted values are hourly averages. The QBER remains well below the \SI{11}{\percent} security threshold (gray dashed line) for this protocol.}
		\label{fig:characterization}
	\end{center}
\end{figure}

\begin{figure*}
	\begin{center}

		\includegraphics[width=\linewidth]{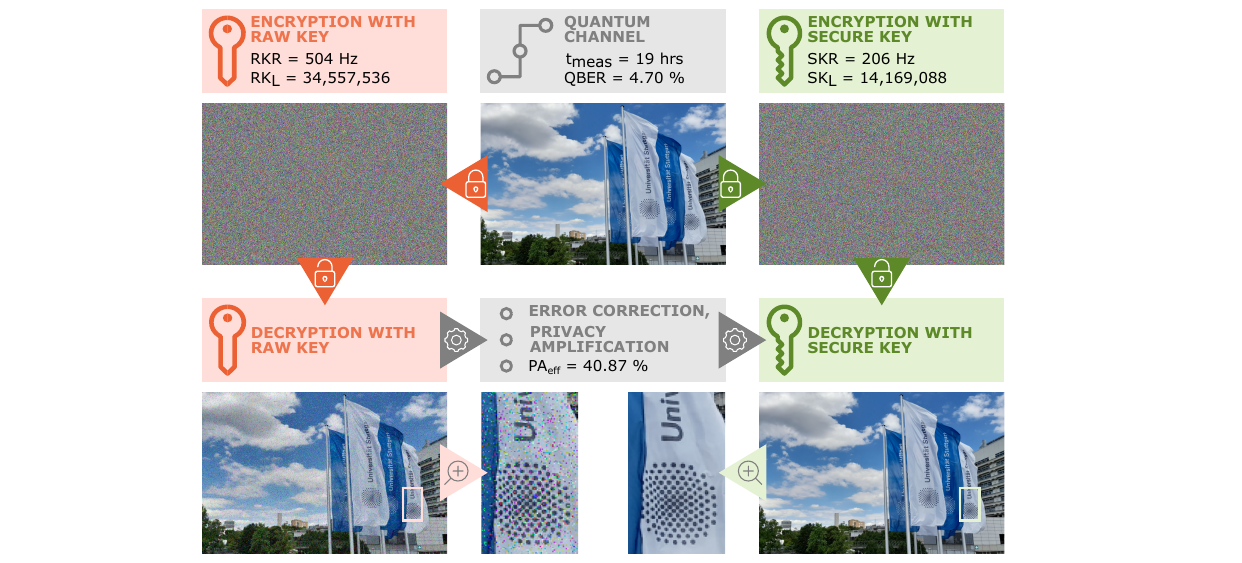}
		\caption{\textbf{Encryption and decryption of a colour image by one-time pad.} (top) The original 24-bit colour image with a resolution of $943\times620$ pixels (shown under the gray box indicating the quantum-channel characteristics, including the measurement time and QBER) is encrypted with the raw key (left side, orange) or with the secure key (right side, green). Decryption using the raw key (bottom left) or the secure key (bottom right), after error correction and privacy amplification with an effective retention ratio of $\text{PA}_{eff}$=\SI{40.48}{\percent}. For both decrypted images, a zoom of the same respective areas highlights the removal of corrupted pixels after decryption with the secure key.}
		\label{fig:fig3}
	\end{center}
\end{figure*}

To mitigate the security threat posed by multiphoton emission, we implemented a squashing model in which Bob assigns a randomly selected measurement outcome corresponding to a single-click event whenever both detectors click~\cite{Gittsovich2014}. This approach increases the QBER as the multiphoton-emission probability increases. In the first step, we optimized the coincidence window between the two photons detection events to maximize the secure key rate (SKR) of the implemented QKD protocol, see Fig.~\ref{fig:characterization}a). Increasing the coincidence window results in a higher number of detected photon pairs and, consequently, a higher raw key rate (RKR), as a larger fraction of photon pairs emitted within the finite radiative lifetime of the QD is captured. However, in contrast, the QBER also increases steadily as the fidelity to the initial Bell state $\ket{\phi^+}$ decreases due to FSS-induced state oscillations and decoherence. In addition, a larger contribution from uncorrelated continuous-wave noise generated by the QFC leads to more false coincidences and increases the probability of both detectors clicking, thereby increasing the frequency of squashing events. The SKR can be calculated as $SKR=RKR\left(1-H_2(QBER)\right)-leak_{EC}$, where $H_2(x)=-x\log_2(x)-(1-x)\log_2(1-x)$ is the binary Shannon entropy and $leak_{EC}$ is the rate of bits leaked during error correction~\cite{Shor2000,Tomamichel2012}. Therefore, the optimal coincidence window for maximizing the SKR represents a trade-off between the increasing raw key rate and the increasing QBER.

To estimate $leak_{EC}$, we implemented an error-correction code based on the low-density parity-check (LDPC) algorithm, in which Alice and Bob share a parity-check matrix. Alice sends the syndrome of her raw key over the public channel, allowing Bob to correct mismatched bits of his noisy copy of the key, at the cost of leaking a known amount, $leak_{EC}$, of information to the public channel that is subsequently suppressed during privacy amplification.~\cite{Kiktenko2017,Rodimin2019,Tomamichel2012}. The raw key was split into finite blocks for error correction, with a block length of \SI{5000}{bits}, determined from optimization tests aimed at minimizing $leak_{EC}$ for QBER values between \SI{4.5}{\percent} and \SI{7.0}{\percent}, see Supplementary Material. For each coincidence window shown in Fig.~\ref{fig:characterization}a), we extracted a \SI{5000}{bit} raw-key block, performed error correction, and calculated the corresponding SKR. The resulting SKR reaches a maximum of \SI{204}{\hertz} at a coincidence window of \SI{275}{\pico\second}.

The QKD protocol was realized over an \SI{19}{\hour}-long measurement period, during which the natural polarization drift of the deployed fibre was found to be minor (see the Supplementary Material). Therefore, the polarization drift was monitored only once per hour, when Alice sent either $H$- or $D$-polarized CW alignment light to Bob, who measured the corresponding Stokes parameter. If this parameter was lower than the threshold value of \SI{0.985}{}, the polarization compensation algorithm was initiated to restore the Stokes parameter to a value above \SI{0.995}{}.

The long-term stability of the raw key rate (RKR) and QBER for the optimal coincidence window is shown in Fig.~\ref{fig:characterization}b). The gradual decrease in the RKR can be attributed to small changes in the wavelength of the converted photons, resulting in their increased loss in a volume Bragg grating (VBG)-based spectral filter with a bandwidth of \SI{25}{\giga\hertz}: this is placed directly after the QFC unit to filter out the spectrally broad noise generated during the conversion process. These small changes in the photon wavelength are caused mostly by temperature fluctuations of the conversion crystal and, on a small part, of the quantum-dot cryostat. Despite that, the QBER shows good stability throughout the entire measurement period, demonstrating the high stability of the deployed setup.

The average RKR over the measurement period was \SI{504}{\hertz}, and the total length of the raw key was over \SI{34.5}{\mega\bit}. The entire raw-key string was divided into \SI{5000}{bit}-long blocks, on which error correction was performed. The SKR was calculated for each block based on the measured QBER and the corresponding number of leaked bits, $leak_{EC}$. The raw-key string was then converted into a secure-key string through privacy amplification using a standard Toeplitz-matrix hash function; see the Supplementary Information for further details. This resulted in an average SKR of \SI{206}{\hertz}, with a total secure-key length of \SI{14.2}{\mega\bit}, corresponding to an overall efficiency of error correction and privacy amplification of \SI{40.87}{\percent} at an average QBER of \SI{4.7}{\percent}.

To further benchmark the protocol and demonstrate its practical usability, the exchanged key was employed to encrypt and decrypt a 24-bit colour image with a resolution of $943\times620$ pixels using a one-time pad (see Fig.~\ref{fig:fig3}, top centre). On the left side (orange pattern), the raw key was used for encryption and decryption, whereas on the right side (green pattern), the secure key obtained after error correction and privacy amplification was used. A closer inspection of the recovered images shows that the error-correction process successfully removed all protocol errors. Further experimental parameters and results are reported in the figure.

The present results demonstrate the maturity of quantum-dot-based sources for hybrid, field-deployed fibre networks incorporating quantum interconnects, such as quantum frequency converters. Effective time synchronization between the two distant locations was achieved by distributing an optical frequency reference through the network, enabling a timing uncertainty below \SI{25}{\pico\second}. Classical communication was established using commercial LAN equipment, demonstrating the possibility of integrating quantum channels with existing classical communication infrastructure. Over a measurement period of \SI{19}{\hour}, an average secure key rate of \SI{206}{\hertz} was achieved, resulting in a total secure-key length of \SI{14.2}{\mega\bit}. This performance was enabled by the bright polarization-entangled photon-pair source and its operation at a repetition rate of \SI{1}{\giga\hertz}. Together, these results demonstrate the feasibility of operating an entanglement-based QKD system continuously in a field-deployed environment while combining quantum-dot photon sources with quantum frequency conversion.

Interestingly, even brighter sources of entangled photons have already been demonstrated, both at shorter wavelengths~\cite{Rota2024}, as in the present study, and directly in the telecommunication C-band~\cite{Joos2026}. These advances provide a pathway towards higher key rates and longer-distance deployed experiments, as well as towards more advanced quantum-network protocols. Together with the demonstrated hybrid quantum interface and deployed-network operation, they represent an important step towards the realization of future large-scale quantum communication networks and, ultimately, a global quantum internet.

\section*{Acknowledgments}

The authors acknowledge K.V.~Reddy from PriTel, Inc. for providing fast photodiodes used in the time synchronization protocol. M. V., R. J., P. V., M. J., S. L. P. and P. M. acknowledge funding by the German Federal Ministry of Research, Technology and Space (BMFTR) via project QR.X (16KISQ013) and QR.N (16KIS2207). Additional funding was also provided via the project EQSOTIC. This project was funded within the QuantERA II Programme that has received funding from the EU’s H2020 research and innovation programme under the GA No 101017733, and with funding organization BMFTR (with project number 16KIS2060K).

\section*{Supplementary material}

See Supplementary Material for a detailed description of the experimental setup and the data processing methods.

\section*{Data availability statement}

The data that support the findings of this study are available from the
corresponding author upon reasonable request.

\bibliography{References}

@article{Weber2019,
  title={Two-photon interference in the telecom C-band after frequency conversion of photons from remote quantum emitters},
  author={Weber, Jonas H and Kambs, Benjamin and Kettler, Jan and Kern, Simon and Maisch, Julian and Vural, H{\"u}seyin and Jetter, Michael and Portalupi, Simone L and Becher, Christoph and Michler, Peter},
  journal={Nature nanotechnology},
  volume={14},
  number={1},
  pages={23--26},
  year={2019},
  publisher={Nature Publishing Group UK London},
  url  = {https://www.nature.com/articles/s41565-018-0279-8}
}

@article{Strobel2024,
  title={High-fidelity distribution of triggered polarization-entangled telecom photons via a 36 km intra-city fiber network},
  author={Strobel, Tim and Kazmaier, Stefan and Bauer, Tobias and Sch{\"a}fer, Marlon and Choudhary, Ankita and Lal Sharma, Nand and Joos, Raphael and Nawrath, Cornelius and Weber, Jonas H and Nie, Weijie and others},
  journal={Optica Quantum},
  volume={2},
  number={4},
  pages={274--281},
  year={2024},
  publisher={Optica Publishing Group},
  url = {https://opg.optica.org/opticaq/abstract.cfm?URI=opticaq-2-4-274}
}

@article{Bozzio2022,
author = {Bozzio, Mathieu and Vyvlecka, Michal and Cosacchi, Michael and Nawrath, Cornelius and Seidelmann, Tim and Loredo, Juan C and Portalupi, Simone L and Axt, Vollrath M and Michler, Peter and Walther, Philip},
doi = {10.1038/s41534-022-00626-z},
journal = {npj Quantum Inf.},
number = {1},
pages = {104},
title = {{Enhancing quantum cryptography with quantum dot single-photon sources}},
url = {https://doi.org/10.1038/s41534-022-00626-z},
volume = {8},
year = {2022}
}

@article{Vyvlecka2023,
    author = {Vyvlecka, Michal and Jehle, Lennart and Nawrath, Cornelius and Giorgino, Francesco and Bozzio, Mathieu and Sittig, Robert and Jetter, Michael and Portalupi, Simone L. and Michler, Peter and Walther, Philip},
    title = "{Robust excitation of C-band quantum dots for quantum communication}",
    journal = {Applied Physics Letters},
    volume = {123},
    number = {17},
    pages = {174001},
    year = {2023},
    month = {10},
    doi = {10.1063/5.0166285},
    url = {https://doi.org/10.1063/5.0166285}
}

@article{zahidy2024,
  title={Quantum key distribution using deterministic single-photon sources over a field-installed fibre link},
  author={Zahidy, Mujtaba and Mikkelsen, Mikkel T and M{\"u}ller, Ronny and Da Lio, Beatrice and Krehbiel, Martin and Wang, Ying and Bart, Nikolai and Wieck, Andreas D and Ludwig, Arne and Galili, Michael and others},
  journal={npj Quantum Information},
  volume={10},
  number={1},
  pages={2},
  year={2024},
  publisher={Nature Publishing Group UK London},
  url = {https://www.nature.com/articles/s41534-023-00800-x}
}

@InProceedings{Bennett1984,
  Title                    = {Quantum cryptography: Public key distribution and coin tossing},
  Author                   = {C. H. Bennett and G. Brassard},
  Booktitle                = {Proceedings of IEEE International Conference on Computers, Systems and Signal Processing},
  Year                     = {1984},
  Pages                    = {175-179}
}

@article{Bozzio2025,
  title = {Quantum cryptography beyond key distribution: Theory and experiment},
  volume = {97},
  ISSN = {1539-0756},
  url = {http://dx.doi.org/10.1103/p84v-1xqv},
  DOI = {10.1103/p84v-1xqv},
  number = {4},
  journal = {Reviews of Modern Physics},
  publisher = {American Physical Society (APS)},
  author = {Bozzio,  Mathieu and Crépeau,  Claude and Wallden,  Petros and Walther,  Philip},
  year = {2025},
  month = Dec 
}

@article{Pan:RevMod20,
	Author = {Xu, Feihu and Ma, Xiongfeng and Zhang, Qiang and Lo, Hoi-Kwong and Pan, Jian-Wei},
	Doi = {10.1103/RevModPhys.92.025002},
	Issue = {2},
	Journal = {Rev. Mod. Phys.},
	Numpages = {60},
	Pages = {025002},
	Publisher = {American Physical Society},
	Title = {Secure quantum key distribution with realistic devices},
	Url = {https://link.aps.org/doi/10.1103/RevModPhys.92.025002},
	Volume = {92},
	Year = {2020}}

@article{NJC:NatComms12,
    doi = {10.1038/ncomms2268},
  url = {https://doi.org/10.1038/ncomms2268},
  year = {2012},
  publisher = {Springer Science and Business Media {LLC}},
  volume = {3},
  number = {1},
  author = {Nelly Huei Ying Ng and Siddarth K. Joshi and Chia Chen Ming and Christian Kurtsiefer and Stephanie Wehner},
  title = {Experimental implementation of bit commitment in the noisy-storage model},
  journal = {Nature Communications}
}

@article{Pan:PRL14,
	Author = {Liu, Yang and Cao, Yuan and Curty, Marcos and Liao, Sheng-Kai and Wang, Jian and Cui, Ke and Li, Yu-Huai and Lin, Ze-Hong and Sun, Qi-Chao and Li, Dong-Dong and Zhang, Hong-Fei and Zhao, Yong and Chen, Teng-Yun and Peng, Cheng-Zhi and Zhang, Qiang and Cabello, Ad\'an and Pan, Jian-Wei},
	Doi = {10.1103/PhysRevLett.112.010504 },
	Issue = {1},
	Journal = {Phys. Rev. Lett.},
	Numpages = {5},
	Pages = {010504},
	Publisher = {American Physical Society},
	Title = {Experimental Unconditionally Secure Bit Commitment},
	Url = {https://link.aps.org/doi/10.1103/PhysRevLett.112.010504}, 
	Volume = {112},
	Year = {2014}}

@article{GAA:pra18,
	Archiveprefix = {arXiv},
	Author = {J-Y. Guan and J-M. Arrazola and R. Amiri and W. Zhang and H. Li and L. You and Z. Wang and Q. Zhang and J-W. Pan},
	Doi = {10.1103/PhysRevA.97.032338},
	Eprint = {1709.05882},
	Journal = {Phys. Rev. A},
	Pages = {032338},
	Title = {Experimental preparation and verification of quantum money},
	Volume = {97},
	Year = {2018}}

@article{BCK:PRA20,
	Author = {Bozzio, Mathieu and Chabaud, Ulysse and Kerenidis, Iordanis and Diamanti, Eleni},
	Doi = {10.1103/PhysRevA.102.022414 },
	Issue = {2},
	Journal = {Phys. Rev. A},
	Numpages = {15},
	Pages = {022414},
	Publisher = {American Physical Society},
	Title = {Quantum weak coin flipping with a single photon},
	Url = {https://link.aps.org/doi/10.1103/PhysRevA.102.022414},
	Volume = {102},
	Year = {2020}}

@article{WEH:Sci18,
	Author = {Wehner, Stephanie and Elkouss, David and Hanson, Ronald},
	Doi = {10.1126/science.aam9288},
	Elocation-Id = {eaam9288},
	Issn = {0036-8075},
	Journal = {Science},
	Number = {6412},
	Publisher = {American Association for the Advancement of Science},
	Title = {Quantum internet: A vision for the road ahead},
	Volume = {362},
	Year = {2018}}

@article{Kimble:2008,
	doi = {10.1038/nature07127},
	url = {https://doi.org/10.1038%2Fnature07127},
	year = 2008,
	publisher = {Springer Science and Business Media {LLC}
},
	volume = {453},
	number = {7198},
	pages = {1023-1030},
	author = {H. J. Kimble},
	title = {The quantum internet},
	journal = {Nature}
}

@article{Peev2009,
  title = {The SECOQC quantum key distribution network in Vienna},
  volume = {11},
  ISSN = {1367-2630},
  url = {http://dx.doi.org/10.1088/1367-2630/11/7/075001},
  DOI = {10.1088/1367-2630/11/7/075001},
  number = {7},
  journal = {New Journal of Physics},
  publisher = {IOP Publishing},
  author = {Peev,  M and Pacher,  C and Alléaume,  R and Barreiro,  C and Bouda,  J and Boxleitner,  W and Debuisschert,  T and Diamanti,  E and Dianati,  M and Dynes,  J F and Fasel,  S and Fossier,  S and F\"{u}rst,  M and Gautier,  J-D and Gay,  O and Gisin,  N and Grangier,  P and Happe,  A and Hasani,  Y and Hentschel,  M and H\"{u}bel,  H and Humer,  G and L\"{a}nger,  T and Legré,  M and Lieger,  R and Lodewyck,  J and Lor\"{u}nser,  T and L\"{u}tkenhaus,  N and Marhold,  A and Matyus,  T and Maurhart,  O and Monat,  L and Nauerth,  S and Page,  J-B and Poppe,  A and Querasser,  E and Ribordy,  G and Robyr,  S and Salvail,  L and Sharpe,  A W and Shields,  A J and Stucki,  D and Suda,  M and Tamas,  C and Themel,  T and Thew,  R T and Thoma,  Y and Treiber,  A and Trinkler,  P and Tualle-Brouri,  R and Vannel,  F and Walenta,  N and Weier,  H and Weinfurter,  H and Wimberger,  I and Yuan,  Z L and Zbinden,  H and Zeilinger,  A},
  year = {2009},
  pages = {075001}
}

@article{Chen_npj2021,
  doi = {10.1038/s41534-021-00474-3},
  url = {https://doi.org/10.1038/s41534-021-00474-3},
  year = {2021},
  publisher = {Springer Science and Business Media {LLC}},
  volume = {7},
  number = {1},
  author = {Teng-Yun Chen and Xiao Jiang and Shi-Biao Tang and Lei Zhou and Xiao Yuan and Hongyi Zhou and Jian Wang and Yang Liu and Luo-Kan Chen and Wei-Yue Liu and Hong-Fei Zhang and Ke Cui and Hao Liang and Xiao-Gang Li and Yingqiu Mao and Liu-Jun Wang and Si-Bo Feng and Qing Chen and Qiang Zhang and Li Li and Nai-Le Liu and Cheng-Zhi Peng and Xiongfeng Ma and Yong Zhao and Jian-Wei Pan},
  title = {Implementation of a 46-node quantum metropolitan area network},
  journal = {npj Quantum Information}
}

@article{SKB:JO19,
	doi = {10.1088/2040-8986/ab05a8},
	url = {https://doi.org/10.1088/2040-8986/ab05a8},
	year = 2019,
	publisher = {{IOP} Publishing},
	volume = {21},
	number = {4},
	pages = {043501},
	journal = {J. Opt.},
	author = {James Schneeloch and Samuel H Knarr and Daniela F Bogorin and Mackenzie L Levangie and Christopher C Tison and Rebecca Frank and Gregory A Howland and Michael L Fanto and Paul M Alsing},
	title = {Introduction to the absolute brightness and number statistics in spontaneous parametric down-conversion}}

@article{BLM:PRL00,
	Author = {Brassard, Gilles and L\"utkenhaus, Norbert and Mor, Tal and Sanders, Barry C.},
	Doi = {10.1103/PhysRevLett.85.1330},
	Issue = {6},
	Journal = {Phys. Rev. Lett.},
	Numpages = {0},
	Pages = {1330--1333},
	Publisher = {American Physical Society},
	Title = {Limitations on Practical Quantum Cryptography},
	Url = {https://link.aps.org/doi/10.1103/PhysRevLett.85.1330},
	Volume = {85},
	Year = {2000}}

@article{Ding2025,
  title = {High-efficiency single-photon source above the loss-tolerant threshold for efficient linear optical quantum computing},
  volume = {19},
  ISSN = {1749-4893},
  url = {http://dx.doi.org/10.1038/s41566-025-01639-8},
  DOI = {10.1038/s41566-025-01639-8},
  number = {4},
  journal = {Nature Photonics},
  publisher = {Springer Science and Business Media LLC},
  author = {Ding,  Xing and Guo,  Yong-Peng and Xu,  Mo-Chi and Liu,  Run-Ze and Zou,  Geng-Yan and Zhao,  Jun-Yi and Ge,  Zhen-Xuan and Zhang,  Qi-Hang and Liu,  Hua-Liang and Wang,  Lin-Jun and Chen,  Ming-Cheng and Wang,  Hui and He,  Yu-Ming and Huo,  Yong-Heng and Lu,  Chao-Yang and Pan,  Jian-Wei},
  year = {2025},
  month = Feb,
  pages = {387–391}
}

@article{Karli2024,
  title = {Controlling the photon number coherence of solid-state quantum light sources for quantum cryptography},
  volume = {10},
  ISSN = {2056-6387},
  url = {http://dx.doi.org/10.1038/s41534-024-00811-2},
  DOI = {10.1038/s41534-024-00811-2},
  number = {1},
  journal = {npj Quantum Information},
  publisher = {Springer Science and Business Media LLC},
  author = {Karli,  Yusuf and Vajner,  Daniel A. and Kappe,  Florian and Hagen,  Paul C. A. and Hansen,  Lena M. and Schwarz,  René and Bracht,  Thomas K. and Schimpf,  Christian and Covre da Silva,  Saimon F. and Walther,  Philip and Rastelli,  Armando and Axt,  Vollrath Martin and Loredo,  Juan C. and Remesh,  Vikas and Heindel,  Tobias and Reiter,  Doris E. and Weihs,  Gregor},
  year = {2024},
  month = Jan 
}

@article{Huber2017,
  title = {Highly indistinguishable and strongly entangled photons from symmetric GaAs quantum dots},
  volume = {8},
  ISSN = {2041-1723},
  url = {http://dx.doi.org/10.1038/ncomms15506},
  DOI = {10.1038/ncomms15506},
  number = {1},
  journal = {Nature Communications},
  publisher = {Springer Science and Business Media LLC},
  author = {Huber,  Daniel and Reindl,  Marcus and Huo,  Yongheng and Huang,  Huiying and Wildmann,  Johannes S. and Schmidt,  Oliver G. and Rastelli,  Armando and Trotta,  Rinaldo},
  year = {2017},
}

@Article{Mueller2014,
author={M{\"u}ller, M.
and Bounouar, S.
and J{\"o}ns, K. D.
and Gl{\"a}ssl, M.
and Michler, P.},
title={On-demand generation of indistinguishable polarization-entangled photon pairs},
journal={Nat. Photonics},
year={2014},
day={16},
publisher={Nature Publishing Group SN  -},
volume={8},
pages={224},
url={http://dx.doi.org/10.1038/nphoton.2013.377}
}

@article{Wei2022,
  title = {Towards Real‐World Quantum Networks: A Review},
  volume = {16},
  ISSN = {1863-8899},
  url = {http://dx.doi.org/10.1002/lpor.202100219},
  DOI = {10.1002/lpor.202100219},
  number = {3},
  journal = {Laser and Photonics Reviews},
  publisher = {Wiley},
  author = {Wei,  Shi‐Hai and Jing,  Bo and Zhang,  Xue‐Ying and Liao,  Jin‐Yu and Yuan,  Chen‐Zhi and Fan,  Bo‐Yu and Lyu,  Chen and Zhou,  Dian‐Li and Wang,  You and Deng,  Guang‐Wei and Song,  Hai‐Zhi and Oblak,  Daniel and Guo,  Guang‐Can and Zhou,  Qiang},
  year = {2022},
  month = Jan 
}

@article{Awschalom2021,
  title = {Development of Quantum Interconnects (QuICs) for Next-Generation Information Technologies},
  volume = {2},
  ISSN = {2691-3399},
  url = {http://dx.doi.org/10.1103/PRXQuantum.2.017002},
  DOI = {10.1103/prxquantum.2.017002},
  number = {1},
  journal = {PRX Quantum},
  publisher = {American Physical Society (APS)},
  author = {Awschalom,  David and Berggren,  Karl K. and Bernien,  Hannes and Bhave,  Sunil and Carr,  Lincoln D. and Davids,  Paul and Economou,  Sophia E. and Englund,  Dirk and Faraon,  Andrei and Fejer,  Martin and Guha,  Saikat and Gustafsson,  Martin V. and Hu,  Evelyn and Jiang,  Liang and Kim,  Jungsang and Korzh,  Boris and Kumar,  Prem and Kwiat,  Paul G. and Lončar,  Marko and Lukin,  Mikhail D. and Miller,  David A.B. and Monroe,  Christopher and Nam,  Sae Woo and Narang,  Prineha and Orcutt,  Jason S. and Raymer,  Michael G. and Safavi-Naeini,  Amir H. and Spiropulu,  Maria and Srinivasan,  Kartik and Sun,  Shuo and Vučković,  Jelena and Waks,  Edo and Walsworth,  Ronald and Weiner,  Andrew M. and Zhang,  Zheshen},
  year = {2021},
}

@article{Couteau2018,
  title = {Spontaneous parametric down-conversion},
  volume = {59},
  ISSN = {1366-5812},
  url = {http://dx.doi.org/10.1080/00107514.2018.1488463},
  DOI = {10.1080/00107514.2018.1488463},
  number = {3},
  journal = {Contemporary Physics},
  publisher = {Informa UK Limited},
  author = {Couteau,  Christophe},
  year = {2018},
  pages = {291–304}
}

@article{Yang2024,
  title = {Tunable quantum dots in monolithic Fabry-Perot microcavities for high-performance single-photon sources},
  volume = {13},
  ISSN = {2047-7538},
  url = {http://dx.doi.org/10.1038/s41377-024-01384-7},
  DOI = {10.1038/s41377-024-01384-7},
  number = {1},
  journal = {Light: Science and Applications},
  publisher = {Springer Science and Business Media LLC},
  author = {Yang,  Jiawei and Chen,  Yan and Rao,  Zhixuan and Zheng,  Ziyang and Song,  Changkun and Chen,  Yujie and Xiong,  Kaili and Chen,  Pingxing and Zhang,  Chaofan and Wu,  Wei and Yu,  Ying and Yu,  Siyuan},
  year = {2024}
}

@article{vanLeent2022,
  title = {Entangling single atoms over 33 km telecom fibre},
  volume = {607},
  ISSN = {1476-4687},
  url = {http://dx.doi.org/10.1038/s41586-022-04764-4},
  DOI = {10.1038/s41586-022-04764-4},
  number = {7917},
  journal = {Nature},
  publisher = {Springer Science and Business Media LLC},
  author = {van Leent,  Tim and Bock,  Matthias and Fertig,  Florian and Garthoff,  Robert and Eppelt,  Sebastian and Zhou,  Yiru and Malik,  Pooja and Seubert,  Matthias and Bauer,  Tobias and Rosenfeld,  Wenjamin and Zhang,  Wei and Becher,  Christoph and Weinfurter,  Harald},
  year = {2022},
  month = July,
  pages = {69–73}
}

@article{Arensktter2023,
  title = {Telecom quantum photonic interface for a 40Ca+ single-ion quantum memory},
  volume = {9},
  ISSN = {2056-6387},
  url = {http://dx.doi.org/10.1038/s41534-023-00701-z},
  DOI = {10.1038/s41534-023-00701-z},
  number = {1},
  journal = {npj Quantum Information},
  publisher = {Springer Science and Business Media LLC},
  author = {Arensk\"{o}tter,  Elena and Bauer,  Tobias and Kucera,  Stephan and Bock,  Matthias and Eschner,  J\"{u}rgen and Becher,  Christoph},
  year = {2023},
}

@article{Bennett1992,
  title = {Quantum cryptography without Bell’s theorem},
  volume = {68},
  ISSN = {0031-9007},
  url = {http://dx.doi.org/10.1103/PhysRevLett.68.557},
  DOI = {10.1103/physrevlett.68.557},
  number = {5},
  journal = {Physical Review Letters},
  publisher = {American Physical Society (APS)},
  author = {Bennett,  Charles H. and Brassard,  Gilles and Mermin,  N. David},
  year = {1992},
  month = Feb,
  pages = {557–559}
}

@article{Rota2024,
   author = {Michele B. Rota and Tobias M. Krieger and Quirin Buchinger and Mattia Beccaceci and Julia Neuwirth and Hêlio Huet and Nikola Horová and Gabriele Lovicu and Giuseppe Ronco and Saimon F. Covre da Silva and Giorgio Pettinari and Magdalena Moczała-Dusanowska and Christoph Kohlberger and Santanu Manna and Sandra Stroj and Julia Freund and Xueyong Yuan and Christian Schneider and Miroslav Ježek and Sven Höfling and Francesco Basso Basset and Tobias Huber-Loyola and Armando Rastelli and Rinaldo Trotta},
   doi = {10.1186/s43593-024-00072-8},
   issn = {2097-1710},
   issue = {1},
   journal = {eLight},
   month = {12},
   pages = {13},
   title = {A source of entangled photons based on a cavity-enhanced and strain-tuned GaAs quantum dot},
   volume = {4},
   url = {https://elight.springeropen.com/articles/10.1186/s43593-024-00072-8},
   year = {2024}
}

@article{Joos2026,
   author = {Raphael Joos and Mohamed Helal and Michal Vyvlečka and Lena Engel and Aditi Javali and Benjamin Breiholz and Ponraj Vijayan and Johannes Michl and Michael Jetter and Sven Höfling and Tobias Huber‐Loyola and Giora Peniakov and Simone L. Portalupi and Peter Michler},
   doi = {10.1002/qute.70317},
   issn = {2511-9044},
   issue = {5},
   journal = {Advanced Quantum Technologies},
   month = {5},
   pages = {e70317},
   publisher = {John Wiley and Sons Inc},
   title = {Telecom C‐Band Source of Triggered Polarization‐Entangled Photon Pairs Combining High Coincidence Rates, Fidelities and Single‐Photon Purity},
   volume = {9},
   url = {https://advanced.onlinelibrary.wiley.com/doi/10.1002/qute.70317},
   year = {2026}
}

@article{Kiktenko2017,
  title = {Symmetric Blind Information Reconciliation for Quantum Key Distribution},
  volume = {8},
  ISSN = {2331-7019},
  url = {http://dx.doi.org/10.1103/physrevapplied.8.044017},
  DOI = {10.1103/physrevapplied.8.044017},
  number = {4},
  journal = {Physical Review Applied},
  publisher = {American Physical Society (APS)},
  author = {Kiktenko,  E. O. and Trushechkin,  A. S. and Lim,  C. C. W. and Kurochkin,  Y. V. and Fedorov,  A. K.},
  year = {2017},
  month = Oct 
}

@article{Rodimin2019,
  title = {Modular Quantum Key Distribution Setup for Research and Development Applications},
  volume = {40},
  ISSN = {1573-8760},
  url = {http://dx.doi.org/10.1007/s10946-019-09793-5},
  DOI = {10.1007/s10946-019-09793-5},
  number = {3},
  journal = {Journal of Russian Laser Research},
  publisher = {Springer Science and Business Media LLC},
  author = {Rodimin,  V. E. and Kiktenko,  E. O. and Usova,  V. V. and Ponomarev,  M. Y. and Kazieva,  T. V. and Miller,  A. V. and Sokolov,  A. S. and Kanapin,  A. A. and Losev,  A. V. and Trushechkin,  A. S. and Anufriev,  M. N. and Pozhar,  N. O. and Kurochkin,  V. L. and Kurochkin,  Y. V. and Fedorov,  A. K.},
  year = {2019},
  month = May,
  pages = {221–229}
}

@article{Tomamichel2012,
  title = {Tight finite-key analysis for quantum cryptography},
  volume = {3},
  ISSN = {2041-1723},
  url = {http://dx.doi.org/10.1038/ncomms1631},
  DOI = {10.1038/ncomms1631},
  number = {1},
  journal = {Nature Communications},
  publisher = {Springer Science and Business Media LLC},
  author = {Tomamichel,  Marco and Lim,  Charles Ci Wen and Gisin,  Nicolas and Renner,  Renato},
  year = {2012},
  month = Jan 
}

@article{Schimpf2021,
  title = {Quantum cryptography with highly entangled photons from semiconductor quantum dots},
  volume = {7},
  ISSN = {2375-2548},
  url = {http://dx.doi.org/10.1126/sciadv.abe8905},
  DOI = {10.1126/sciadv.abe8905},
  number = {16},
  journal = {Science Advances},
  publisher = {American Association for the Advancement of Science (AAAS)},
  author = {Schimpf,  Christian and Reindl,  Marcus and Huber,  Daniel and Lehner,  Barbara and Covre Da Silva,  Saimon F. and Manna,  Santanu and Vyvlecka,  Michal and Walther,  Philip and Rastelli,  Armando},
  year = {2021},
  month = Apr 
}

@book{Michler2017,
   city = {Cham},
   doi = {10.1007/978-3-319-56378-7},
   editor = {Peter Michler},
   isbn = {978-3-319-56377-0},
   month = {7},
   publisher = {Springer International Publishing},
   title = {Quantum Dots for Quantum Information Technologies},
   url = {http://link.springer.com/10.1007/978-3-319-56378-7},
   year = {2017}
}

@article{Patel2010,
   author = {Raj B. Patel and Anthony J. Bennett and Ian Farrer and Christine A. Nicoll and David A. Ritchie and Andrew J. Shields},
   doi = {10.1038/nphoton.2010.161},
   issn = {1749-4885},
   issue = {9},
   journal = {Nature Photonics},
   month = {9},
   pages = {632-635},
   publisher = {Nature Publishing Group},
   title = {Two-photon interference of the emission from electrically tunable remote quantum dots},
   volume = {4},
   url = {https://www.nature.com/articles/nphoton.2010.161},
   year = {2010}
}

@article{Martin-Sanchez2018,
   author = {Javier Martín-Sánchez and Rinaldo Trotta and Antonio Mariscal and Rosalía Serna and Giovanni Piredda and Sandra Stroj and Johannes Edlinger and Christian Schimpf and Johannes Aberl and Thomas Lettner and Johannes Wildmann and Huiying Huang and Xueyong Yuan and Dorian Ziss and Julian Stangl and Armando Rastelli},
   doi = {10.1088/1361-6641/aa9b53},
   issn = {0268-1242},
   issue = {1},
   journal = {Semiconductor Science and Technology},
   month = {1},
   pages = {013001},
   publisher = {IOP Publishing},
   title = {Strain-tuning of the optical properties of semiconductor nanomaterials by integration onto piezoelectric actuators},
   volume = {33},
   url = {http://stacks.iop.org/0268-1242/33/i=1/a=013001?key=crossref.9511d6c97e631ba66726f03c8bfa0f38},
   year = {2018}
}

@article{Zhang2025,
  title = {Experimental Single-Photon Quantum Key Distribution Surpassing the Fundamental Weak Coherent-State Rate Limit},
  volume = {134},
  ISSN = {1079-7114},
  url = {http://dx.doi.org/10.1103/PhysRevLett.134.210801},
  DOI = {10.1103/physrevlett.134.210801},
  number = {21},
  journal = {Physical Review Letters},
  publisher = {American Physical Society (APS)},
  author = {Zhang,  Yang and Ding,  Xing and Li,  Yang and Zhang,  Likang and Guo,  Yong-Peng and Wang,  Gao-Qiang and Ning,  Zhen and Xu,  Mo-Chi and Liu,  Run-Ze and Zhao,  Jun-Yi and Zou,  Geng-Yan and Wang,  Hui and Cao,  Yuan and He,  Yu-Ming and Peng,  Cheng-Zhi and Huo,  Yong-Heng and Liao,  Sheng-Kai and Lu,  Chao-Yang and Xu,  Feihu and Pan,  Jian-Wei},
  year = {2025},
  month = May 
}

@article{BassoBasset2021,
  title = {Quantum key distribution with entangled photons generated on demand by a quantum dot},
  volume = {7},
  ISSN = {2375-2548},
  url = {http://dx.doi.org/10.1126/sciadv.abe6379},
  DOI = {10.1126/sciadv.abe6379},
  number = {12},
  journal = {Science Advances},
  publisher = {American Association for the Advancement of Science (AAAS)},
  author = {Basso Basset,  Francesco and Valeri,  Mauro and Roccia,  Emanuele and Muredda,  Valerio and Poderini,  Davide and Neuwirth,  Julia and Spagnolo,  Nicolò and Rota,  Michele B. and Carvacho,  Gonzalo and Sciarrino,  Fabio and Trotta,  Rinaldo},
  year = {2021},
  month = Mar 
}

@article{Gyger2022,
  title = {Metropolitan single-photon distribution at 1550 nm for random number generation},
  volume = {121},
  ISSN = {1077-3118},
  url = {http://dx.doi.org/10.1063/5.0112939},
  DOI = {10.1063/5.0112939},
  number = {19},
  journal = {Applied Physics Letters},
  publisher = {AIP Publishing},
  author = {Gyger,  Samuel and Zeuner,  Katharina D. and Lettner,  Thomas and Bensoussan,  Sandra and Carln\"{a}s,  Martin and Ekemar,  Liselott and Schweickert,  Lucas and Hedlund,  Carl Reuterski\"{o}ld and Hammar,  Mattias and Nilsson,  Tigge and Alml\"{o}f,  Jonas and Steinhauer,  Stephan and Llosera,  Gemma Vall and Zwiller,  Val},
  year = {2022},
  month = Nov 
}

@article{BrookeBarnes2026,
  title = {Decoy-State Quantum Key Distribution over 227 km with a Frequency-Converted Telecom Single-Photon Source},
  volume = {137},
  ISSN = {1079-7114},
  url = {http://dx.doi.org/10.1103/5wwp-rbpm},
  DOI = {10.1103/5wwp-rbpm},
  number = {6},
  journal = {Physical Review Letters},
  publisher = {American Physical Society (APS)},
  author = {Brooke Barnes,  Frederik and Pousa,  Roberto G. and Morrison,  Christopher L. and Koong,  Zhe Xian and Ho,  Joseph and Graffitti,  Francesco and Jeffers,  John and Oi,  Daniel K. L. and Gerardot,  Brian D. and Fedrizzi,  Alessandro},
  year = {2026},
  month = Aug 
}

@article{Vajner2026,
  title = {Single-photon advantage in quantum cryptography beyond QKD},
  volume = {17},
  ISSN = {2041-1723},
  url = {http://dx.doi.org/10.1038/s41467-026-69995-9},
  DOI = {10.1038/s41467-026-69995-9},
  number = {1},
  journal = {Nature Communications},
  publisher = {Springer Science and Business Media LLC},
  author = {Vajner,  Daniel A. and Kaymazlar,  Koray and Drauschke,  Fenja and Rickert,  Lucas and von Helversen,  Martin and Liu,  Hanqing and Li,  Shulun and Ni,  Haiqiao and Niu,  Zhichuan and Pappa,  Anna and Heindel,  Tobias},
  year = {2026},
  month = Feb 
}

@article{Gittsovich2014,
  title = {Squashing model for detectors and applications to quantum-key-distribution protocols},
  volume = {89},
  ISSN = {1094-1622},
  url = {http://dx.doi.org/10.1103/PhysRevA.89.012325},
  DOI = {10.1103/physreva.89.012325},
  number = {1},
  journal = {Physical Review A},
  publisher = {American Physical Society (APS)},
  author = {Gittsovich,  O. and Beaudry,  N. J. and Narasimhachar,  V. and Alvarez,  R. Romero and Moroder,  T. and L\"{u}tkenhaus,  N.},
  year = {2014},
  month = Jan 
}

@article{Rota2026,
  title = {Quantum key distribution with entanglement-swapped photons},
  ISSN = {2643-1564},
  url = {http://dx.doi.org/10.1103/68f1-cjfk},
  DOI = {10.1103/68f1-cjfk},
  journal = {Physical Review Research},
  publisher = {American Physical Society (APS)},
  author = {Rota,  Michele B. and Basset,  Francesco Basso and Laneve,  Alessandro and Salusti,  Francesco and Claro-Rodriguez,  Nicolas and Ronco,  Giuseppe and Beccaceci,  Mattia and Krieger,  Tobias M. and Buchinger,  Quirin and da Silva,  Saimon F. Covre and Stroj,  Sandra and Gumberidze,  Mariia and Usenko,  Vladyslav and H\"{o}fling,  Sven and Huber-Loyola,  Tobias and Castaneda,  Mario A. Usuga and Rastelli,  Armando and J\"{o}ns,  Klaus D. and Trotta,  Rinaldo},
  year = {2026},
  month = July 
}

@article{Bock2018,
  title = {High-fidelity entanglement between a trapped ion and a telecom photon via quantum frequency conversion},
  volume = {9},
  ISSN = {2041-1723},
  url = {http://dx.doi.org/10.1038/s41467-018-04341-2},
  DOI = {10.1038/s41467-018-04341-2},
  number = {1},
  journal = {Nature Communications},
  publisher = {Springer Science and Business Media LLC},
  author = {Bock,  Matthias and Eich,  Pascal and Kucera,  Stephan and Kreis,  Matthias and Lenhard,  Andreas and Becher,  Christoph and Eschner,  J\"{u}rgen},
  year = {2018},
  month = May 
}

@article{Shor2000,
  title = {Simple Proof of Security of the BB84 Quantum Key Distribution Protocol},
  volume = {85},
  ISSN = {1079-7114},
  url = {http://dx.doi.org/10.1103/PhysRevLett.85.441},
  DOI = {10.1103/physrevlett.85.441},
  number = {2},
  journal = {Physical Review Letters},
  publisher = {American Physical Society (APS)},
  author = {Shor,  Peter W. and Preskill,  John},
  year = {2000},
  pages = {441–444}
}

@article{Ma2012,
  title = {Quantum teleportation over 143 kilometres using active feed-forward},
  volume = {489},
  ISSN = {1476-4687},
  url = {http://dx.doi.org/10.1038/nature11472},
  DOI = {10.1038/nature11472},
  number = {7415},
  journal = {Nature},
  publisher = {Springer Science and Business Media LLC},
  author = {Ma,  Xiao-Song and Herbst,  Thomas and Scheidl,  Thomas and Wang,  Daqing and Kropatschek,  Sebastian and Naylor,  William and Wittmann,  Bernhard and Mech,  Alexandra and Kofler,  Johannes and Anisimova,  Elena and Makarov,  Vadim and Jennewein,  Thomas and Ursin,  Rupert and Zeilinger,  Anton},
  year = {2012},
  pages = {269–273}
}

@article{Ren2017,
  title = {Ground-to-satellite quantum teleportation},
  volume = {549},
  ISSN = {1476-4687},
  url = {http://dx.doi.org/10.1038/nature23675},
  DOI = {10.1038/nature23675},
  number = {7670},
  journal = {Nature},
  publisher = {Springer Science and Business Media LLC},
  author = {Ren,  Ji-Gang and Xu,  Ping and Yong,  Hai-Lin and Zhang,  Liang and Liao,  Sheng-Kai and Yin,  Juan and Liu,  Wei-Yue and Cai,  Wen-Qi and Yang,  Meng and Li,  Li and Yang,  Kui-Xing and Han,  Xuan and Yao,  Yong-Qiang and Li,  Ji and Wu,  Hai-Yan and Wan,  Song and Liu,  Lei and Liu,  Ding-Quan and Kuang,  Yao-Wu and He,  Zhi-Ping and Shang,  Peng and Guo,  Cheng and Zheng,  Ru-Hua and Tian,  Kai and Zhu,  Zhen-Cai and Liu,  Nai-Le and Lu,  Chao-Yang and Shu,  Rong and Chen,  Yu-Ao and Peng,  Cheng-Zhi and Wang,  Jian-Yu and Pan,  Jian-Wei},
  year = {2017},
  month = Aug,
  pages = {70–73}
}

\end{document}